%% file: bmc_article.tex
\documentclass[10pt]{bmc_article}

\usepackage{cite} 
\usepackage{url}  
\usepackage{ifthen}  
\usepackage{multicol}   
\usepackage[utf8]{inputenc} 
 \usepackage{amsmath}
\usepackage{amsfonts}
\usepackage{graphicx}
\usepackage[usenames,dvipsnames]{color}

\newboolean{publ}

\newenvironment{bmcformat}{\begin{raggedright}\baselineskip20pt\sloppy\setboolean{publ}{false}}{\end{raggedright}\baselineskip20pt\sloppy}

\begin{document}
\begin{bmcformat}

\title{Graphical models for inferring single molecule dynamics}

\author{Jonathan E. Bronson\correspondingauthor$^{1}$%
       \email{Jonathan E. Bronson\correspondingauthor\ - jeb2126@columbia.edu}%
       \and
       Jake M. Hofman$^{2}$%
       \email{Jake M. Hofman - jmh2045@columbia.edu}%
      \and
      Jingyi Fei$^1$%
      \email{Jingyi Fei - jf2276@columbia.edu}%
      \and
      Ruben L. Gonzalez, Jr.$^1$ %
      \email{Ruben L. Gonzalez, Jr. - rlg2118@columbia.edu}%
      and
         Chris H. Wiggins$^3$%
         \email{Chris H. Wiggins - chris.wiggins@columbia.edu}%
     }

\address{%
    \iid(1)Department of Chemistry, Columbia University, New York, NY 10027, USA\\
    \iid(2)Yahoo! Research, 111 West 40th St., New York, NY 10018, USA\\
    \iid(3)Department of Applied Physics and Applied Mathematics, Columbia University, %
        New York, NY 10027, USA
}%

\maketitle

\input{aliases}
\begin{abstract}
  \paragraph*{Background:} The recent explosion of experimental
  techniques in single molecule biophysics has generated a variety of
  novel time series data requiring equally novel computational tools
  for analysis and inference.  This article describes in general terms
  how graphical modeling may be used to learn from biophysical time
  series data using the variational Bayesian expectation maximization
  algorithm (VBEM).  The discussion is illustrated by the example of
  single-molecule fluorescence resonance energy transfer (smFRET) {\it
    versus} time data, where the smFRET time series is modeled as a
  hidden Markov model (HMM) with Gaussian observables. A detailed
  description of smFRET is provided as well.
      
  \paragraph*{Results:}The VBEM algorithm returns the model's evidence
  and an approximating posterior parameter distribution given the
  data. The former provides a metric for model selection via maximum
  evidence (ME), and the latter a description of the model's
  parameters learned from the data. ME/VBEM provide several advantages
  over the more commonly used approach of maximum likelihood (ML)
  optimized by the expectation maximization (EM) algorithm, the most
  important being a natural form of model selection and a well-posed
  (non-divergent) optimization problem.
 
  \paragraph*{Conclusions:} The results demonstrate the utility of
  graphical modeling for inference of dynamic processes in single
  molecule biophysics.
\end{abstract}

\ifthenelse{\boolean{publ}}{\begin{multicols}{2}}{}

 \setlength{\parskip}{\baselineskip}

\section*{Background}

 \input{intro.tex}

 \subsection*{Graphical models}
   \label{sec:GM}
   \input{gm.tex}

\subsection*{Inference of GMs}
\label{sec:inference}
\input{inference}

 \subsubsection*{Inference via maximum likelihood}
   \label{sec:ML}
   \input{ml.tex}

 \subsubsection*{Inference via maximum evidence}
   \label{sec:ME}
   \input{me.tex}

 \subsection*{smFRET}
   \label{sec:FRET_intro}
   \input{fret_background.tex}

\section*{Results and Discussion}

 \subsection*{smFRET as a graphical model}
   \label{sec:FRET_model}
   \input{modeling_fret.tex}

 \subsection*{Illustration of the inference}
   \label{sec:validation}
   \input{validation.tex}

\section*{Conclusions}
   \label{sec:conclusion}
   \input{conclusion.tex}

\section*{Methods}

All code used in this manuscript is available open source at
http://vbfret.sourceforge.net/.

\subsection*{Variational Bayesian expectation maximization}
   \label{sec:VBEM}
   \input{vbem.tex}

 \subsection*{Priors}
\input{priors}

\subsection*{Data generation}
 \label{sec:supporting}
 \input{supporting.tex}

\section*{Competing interests}
The authors declare that they have no competing interests.

\section*{Authors contributions}
JEB contributed to the graphical modeling and smFRET inference.
JMH contributed to the graphical modeling.
JF and RLG contributed to the smFRET inference.
CHW contributed to the graphical modeling and smFRET inference.
JEB and CHW wrote the manuscript.

\section*{Acknowledgments}
This work was supported by a grant to CHW from the NIH
(5PN2EY016586-03); grants to RLG from the Burroughs Wellcome Fund
(CABS 1004856), the NSF (MCB 0644262), and the NIH-NIGMS
(1RO1GM084288-01); and a grant to JEB from the NSF (GRFP).

{\ifthenelse{\boolean{publ}}{\footnotesize}{\small}
 \bibliographystyle{bmc_article}  
  \bibliography{bmc_article} }     

\ifthenelse{\boolean{publ}}{\end{multicols}}{}

\input{captions_for_figures.tex}

\end{bmcformat}
\end{document}

%% file: aliases.tex
\newcommand{\jeb}[1]{{\color{ForestGreen}{[JEB: #1]}}}
\newcommand{\chw}[1]{{\color{Red}{[CHW: #1]}}}
\newcommand{\fig}[1]{Fig.~\ref{#1}}
\newcommand{\sect}[1]{Sec.~\ref{#1}}
\newcommand{\eq}[1]{Eq.~\ref{#1}}
\newcommand{\tab}[1]{Table~\ref{#1}}
\newcommand{\ie}{{\it{i.e.}}}
\newcommand{\eg}{{\it{e.g.}}}
\newcommand{\cf}{{\it{cf.}}}
\newcommand{\rev}[1]{{\color{RoyalBlue}{#1}}}

\newcommand{\chk}[1]{{\color{red}{#1}}} 
\newcommand{\chem}[1]{\ensuremath{\mathrm{#1}}}
\renewcommand{\sup}[1]{\ensuremath{^{\mathrm{#1}}}}
\newcommand{\sub}[1]{\ensuremath{_{\mathrm{#1}}}}

\newcommand{\FRET}{{\small{\sc {FRET}}}}
\newcommand{\vbFRET}{{\small{\sc {vbFRET}}}}
\newcommand{\vbf}{ME}
\newcommand{\mlm}{ML}
\newcommand{\kclose}{\ensuremath{k_{\mathrm{close}}}}
\newcommand{\kopen}{\ensuremath{k_{\mathrm{open}}}}
\newcommand{\midval}{$\sim.35$ FRET }
\newcommand{\midvalstate}{$\sim.35$ FRET state }
\newcommand{\flo}{f_{\rm low}}
\newcommand{\fhi}{f_{\rm high}}
\newcommand{\fmid}{f_{\rm mid}}
\newcommand{\fmidstate}{$\fmid$ state}
\newcommand{\matlab}{{\small{\sc {matlab}}}}

\newcommand{\ajk}{\ensuremath{a_{jk}}}
\newcommand{\D}{\ensuremath{{\bf D}}}
\renewcommand{\d}{\ensuremath{d}}
\newcommand{\X}{\ensuremath{{\bf X}}}
\newcommand{\Z}{\ensuremath{{\bf Z}}}
\newcommand{\z}{\ensuremath{z}}
\newcommand{\thv}{\ensuremath{\vec{\theta}}}
\newcommand{\thvs}{\ensuremath{\vec{\theta}_*}}
\newcommand{\thvml}{\ensuremath{\hat{\theta}_{\rm ML}}}
\newcommand{\thvmap}{\ensuremath{\hat{\theta}_{\rm MAP}}}
\newcommand{\pivec}{\ensuremath{\vec{\pi}}}
\newcommand{\m}{\ensuremath{m}}
\newcommand{\K}{\ensuremath{K}}
\newcommand{\Ks}{\ensuremath{K_*}}
\newcommand{\zhat}{{\hat\Z}}
\newcommand{\A}{\ensuremath{A}}

\newcommand{\argmax}{\operatornamewithlimits{argmax}}
\newcommand{\argmin}{\operatornamewithlimits{argmin}}
\newcommand{\ith}{\ensuremath{i^{th}}}
\newcommand{\jth}{\ensuremath{j^{th}}}
\newcommand{\kth}{\ensuremath{k^{th}}}
\newcommand{\nth}{\ensuremath{n^{th}}}
\newcommand{\eqspace}{\vspace{-7.5mm}}
\newcommand{\Dkl}[2]{\ensuremath{D_{KL}\left({#1}||{#2}\right)}}
\newcommand{\dkl}{\ensuremath{D_{KL}}}

\newcommand{\muvec}{\ensuremath{\vec{\mu}}}
\newcommand{\lambdavec}{\ensuremath{\vec{\lambda}}}
\newcommand{\Sigvec}{\ensuremath{\vec{\Sigma}}}
\newcommand{\alphavec}{\ensuremath{\vec{\alpha}}}
\newcommand{\muo}{\ensuremath{\mu_o}}
\newcommand{\varo}{\ensuremath{\sigma^2_o}}

\newcommand{\upik}{\ensuremath{u_{\pi}^k}}
\newcommand{\ua}{\ensuremath{u_{a}^k}}
\newcommand{\uajk}{\ensuremath{u_{a}^{jk}}}
\newcommand{\umk}{\ensuremath{u_{\mu}^{k}}}
\newcommand{\ubk}{\ensuremath{u_{\beta}^{k}}}
\newcommand{\uvk}{\ensuremath{u_{v}^{k}}}
\newcommand{\uwk}{\ensuremath{u_{W}^{k}}}
\newcommand{\bupi}{\ensuremath{{\vec{u}_{\pi}}}}
\newcommand{\bua}{\ensuremath{\vec{u}_{a}}}
\newcommand{\bum}{\ensuremath{\vec{u}_{\mu}}}
\newcommand{\bub}{\ensuremath{\vec{u}_{\beta}}}
\newcommand{\buv}{\ensuremath{\vec{u}_{v}}}
\newcommand{\buW}{\ensuremath{\vec{u}_{W}}}
\newcommand{\uvec}{\ensuremath{\vec{u}}}

\newcommand{\priorm}{\ensuremath{p(\thv|\m)}}
\newcommand{\postm}{\ensuremath{p(\thv|\D,\m)}}
\newcommand{\evm}{\ensuremath{p(\D|\m)}}
\newcommand{\likem}{\ensuremath{p(\D|\thv,\m)}}
\newcommand{\mone}{\ensuremath{\m_1}}
\newcommand{\mtwo}{\ensuremath{\m_2}}
\newcommand{\kone}{\ensuremath{\K_1}}
\newcommand{\ktwo}{\ensuremath{\K_2}}
\newcommand{\prior}{\ensuremath{p(\thv|\uvec,\K)}}
\newcommand{\post}{\ensuremath{p(\thv|\D,\uvec,\K)}}
\newcommand{\ev}{\ensuremath{p(\D|\uvec,\K)}}
\newcommand{\like}{\ensuremath{p(\D|\thv,\K)}}
\newcommand{\npar}{\ensuremath{\mathcal{M}}}
\newcommand{\lb}{\ensuremath{\mathcal{L}(q)}}
\newcommand{\qzt}{\ensuremath{q(\Z,\thv)}}
\newcommand{\qzts}{\ensuremath{q_*(\Z,\thv)}}
\newcommand{\jointpost}{\ensuremath{p(\Z,\thv|\D,\uvec,\K)}}
\newcommand{\jointlike}{\ensuremath{p(\D,\Z|\thv,\K)}}
\newcommand{\zpost}{\ensuremath{p(\Z|\D,\thv,\K)}}
\newcommand{\kopt}{,\K}
\newcommand{\re}[1]{{\rm e}^{#1}}
\newcommand{\infinity}{\infty}
\newcommand{\pyztGk}{\ensuremath{p(\D,\Z,\thv|\K)}}
\newcommand{\bx}{\ensuremath{{\bf x}}}
\newcommand{\bff}{\ensuremath{{\bf f}}}

\renewcommand{\rev}[1]{{{#1}}}

%% file: intro.tex
Single-molecule techniques allow biophysicists to probe the dynamics
of proteins, nucleic acids, and other biological macromolecules with
unprecedented resolution \cite{HaARB08,HaCOSB10,DekkerCOSB07}.  It is
now possible to observe viruses pack DNA into capsids
\cite{BustamanteNAT09}, helicases unzip nucleic acids
\cite{BustamanteNAT06}, motor proteins walk on biopolymers
\cite{TomishigeNAT07}, and ribosome domains undergo structural
rearrangements during translation \cite{GonzalezPNAS09}. These data
are acquired by recording the fluorescent output or forces generated
from, for example, biomolecules tethered onto microscope slides
\cite{GonzalezMC08}; walking on biopolymers \cite{SchnappNAT90};
diffusing in hydrodynamic flow cells \cite{GreeneNSMB09}; or pulled by
optical \cite{ChuSCI94} or magnetic \cite{MarkoPRE04} tweezers.  Often
the molecules studied move through a series of locally stable
molecular conformations or positions (generically termed states) and
give rise to data of the type shown in \fig{fig:time_series}.  From
these data, the experimentalist wishes to learn a model describing the
number of states occupied by the molecule and the transition rates
between states.  Although the myriad experimental techniques available
have much in common, the data they generate often differ enough to
require unique models.

For example, some of these models will involve conversion of chemical
to mechanical energy, or motion associated with diffusion, or motion
associated with transitions between distinct configurational
states. Modeling the data, then, typically involves introducing
several variables --- some of which are observed, others of which are
latent or ``hidden"; some of which are real-valued coordinates, others
of which are discrete states --- and specifying algebraically how they
are related. Such algebraic relations among a few variables are
typical in physical modeling (e.g., the stochastic motion of a random
walker, or the assumption of additive, independent, normally
distributed errors typical in regression); models involving multiple
conditionally-dependent observations or hidden variables with more
structured noise behavior are less common.  Implicitly, each equation
of motion or of constraint specifies which variables are conditionally
dependent and which are conditionally independent.  Graphical
modeling, which begins with charting these dependencies among a set of
nodes, with edges corresponding to the conditional probabilities which
must be algebraically specified (\ie, the typical elements of a
physical model) organizes this process and facilitates
basing inference on such models 
\cite{jordan1999, MacKay03,Bishop06}.

Here we explore the application of a specific subset of GMs to
biophysical time series data using a specific algorithmic approach for
inference: the directed GM and the variational Bayesian expectation
maximization algorithm (VBEM). After discussing the theoretical basis
and practical advantages of this general approach \rev{to modeling
  biophysical time series data}, we \rev{apply the method}
to the problem of inference given single molecule fluorescence
resonance energy transfer (smFRET) time series data. \rev{We emphasize
  the process and caveats of modeling smFRET data with a GM and
  demonstrate the most helpful features of VBEM for this type of time
  series inference.}

%% file: gm.tex
GMs are a flexible inference framework based on factorizing a
(high-dimensional) multivariate joint distribution into
(lower-dimensional) conditionals and marginals
\cite{jordan1999,MacKay03, Bishop06}.  
In a GM, the nodes of the graph represent either
observable variables (data, denoted by filled circles), latent
variables (hidden states, denoted by open circles), or fixed
parameters (denoted by dots). 
Directed edges between nodes represent conditional probabilities.
  \rev{For example, the three-node graphical model $X\rightarrow Y\rightarrow Z$ implies that the joint distribution $p(Z,Y,X)\equiv p(Z|Y,X)p(Y|X)p(X)$ can be further factorized as $p(Z|Y)p(Y|X)p(X)$.}
  \rev{Data with a temporal component are modeled by
  connecting arrows from variables at earlier time steps to variables
  at later time steps.}
In many graphical models, the number of observed and latent variables
grows with the size of the data set under consideration. To avoid
clutter, these variables are written once and placed in a box, often
called a ``plate'', labeled with the number of times the variables are
repeated \cite{Bishop06}. This manuscript will denote hidden variables
by $\z$ and observed data by $\d$. \rev{Parameters which are vectors
  will be denoted as such by 
  overhead arrows}.

As an example of a simple GM, imagine trying to learn the number of
boys and girls in an elementary school class of $N$ students from a
table of their heights and weights. \rev{Here the hidden variable is gender
and the observed variable, (height, weight), is 
a random variable conditionally dependent on 
the hidden variable. 
The resulting GM is shown in \fig{fig:gm_example}, with the
parameters of $p(gender)$ denoted by $\vec{\alpha}$ and the parameters
$p(height,weight|gender)$ denoted by $\vec{\mu}$ and $\vec{\Sigma}$.}
The  expression for the probability of the observed data
($\{\d_1,\ldots,\d_N\} = \D$) and latent genders
($\{\z_1,\ldots,\z_N\} = \Z$) is uniquely specified by the graph
and the factorization it implies:
\begin{equation}
  p(\D,\Z|\muvec,\Sigvec,\alphavec) = \prod_{n=1}^N p(\d_n|\z_n,\muvec,\Sigvec)p(\z_n|\alphavec).
  \label{eq:boy_girl}
\end{equation}
In such a simple case it is straighforward
to arrive at the expression in \eq{eq:boy_girl} without
the use of a GM, but such a chart makes this factorization far more
obvious and interpretable.

%% file: inference.tex
In some contexts, one wishes to learn the probability of the hidden
states given the observed data, \zpost, where \thv\ denotes the
parameters of the model and 
\K\ 
denotes the
number of allowed values of the latent variables (i.e. number of
hidden states). If \thv\ is known then efficient inference of $\zpost$
can be performed on any loop-free graph with discrete latent states
using the {\it sum-product} algorithm \cite{LoeligerIEEE01}, or, if
only the most probable values of \Z\ are needed, using the closely
related {\it max-sum} algorithm \cite{FreemanIEEE01}.  A loop in a
graph occurs when multiple pathways connect two variables, which is
unlikely in a graph modeling time series data.  Inference for models
with continuous latent variables is discussed in
\cite{BealNIPS01,BishopNIPS03}.  \rev{For most time series inference
  problems in biophysics,} both \Z\ and \thv\ are unknown. In these
cases, a criterion for choosing a best estimate of \thv\ and an
optimization 
algorithm
to find this estimate are needed.

%% file: ml.tex
Estimating \thv\ is most commonly accomplished using
the {\it maximum likelihood} (ML) method, which estimates \thv\ 
as
\begin{equation}
  \thvml = \argmax_{\thv} \like = \argmax_{\thv}\sum_\Z \jointlike.
\label{eq:ml}
\end{equation}
The probability \like\ is known as the {\it likelihood}.  The
expectation maximization (EM) algorithm can be used to \rev{estimate}
\thvml\ \cite{Dempster77}. In EM, an initial guess for \thvml\ is used
to calculate \zpost. The \zpost\ learned is then used to calculate a
new guess for \thvml. The algorithm iterates until convergence, and is
guaranteed to converge to a local optimum. The EM algorithm should be
run with multiple initializations of \thvml, often called ``random
restarts'', to 
increase
the probability of finding the globally
optimal \thvml.

ML solved via EM is a generally effective method to perform inference
however, it has two prominent shortcomings\cite{MacKay03,
  Bishop06}:

{\bf Model selection:} The first limitation of ML is that it has no
form of model selection: the likelihood 
monotonically
increases with the addition of more model parameters.  This problem of
fitting too many states to the data (overfitting) is highly
undesirable for biophysical time series data, where learning the
correct \K\ for the data is often an experimental objective.

{\bf Ill-posedness} The second problem with ML occurs only in the case
of a model with multiple hidden states and a continuous observable
(a case which includes the majority of biophysical time series data,
\rev{including the smFRET data we will consider here}). \rev{If the
  mean of one hidden state 
  approaches
  the position of a data
  point and the variance of that state 
  approaches
  zero,
  the contribution of that datum to the likelihood will
  diverge.} When this happens, the likelihood will be infinite
regardless of how poorly the rest of the data are modeled,
\rev{provided the other states in the model have non-zero
  probabilities for the rest of the data.}
For 
such
models, the ML method is ill-posed; \rev{poor
  parameters can still result in infinite likelihood}.

In practical contexts, the second problem (divergent likelihood) can
be avoided either by performing MAP estimation (maximizing the
likelihood times a prior which penalizes small variance) or by
ignoring solutions for which likelihood is diverging.
That is, one does not actually maximize the likelihood. Model
selection can then be argued for based on cross-validation or by
penalizing likelihood with a term which monotonically increases with
model complexity \cite{AIC, schwarz1978edm, Bishop06}. We consider,
instead, an alternative optimization criterion which naturally avoids
these problems.

%% file: me.tex
A Bayesian alternative to ML is to perform inference using the {\it
  maximum evidence} (ME) method. ME can be thought of as an extension
of ML to the problem of model selection. Where ML asks which
parameters maximize the probability of the data for a given model, ME
asks which model, including nested models which differ only in \K,
makes the data most probable. According to ME, the model of correct
complexity (\Ks) is
\begin{equation}
\Ks = \argmax_\K \ev = \argmax_\K \sum_\Z\int d\thv \jointlike \prior.
\label{eq:me}
\end{equation}
The quantity \ev\ is called the evidence. Sometimes it is also
referred to as the marginal likelihood, since unknown parameters are
assigned probability distributions and marginalized (or summed out)
over all possible settings. The evidence penalizes both models which
underfit and models which overfit.  The second expression in
\eq{eq:me} follows readily from the sum rule of probability provided
we are willing to model the parameters themselves as random
variables. That is, we are willing to specify a distribution over
parameters, \prior. This distribution is called the ``prior'', since
it can be thought of as the probability of the parameters prior to
seeing any data. The parameters for the distributions of the prior
(\uvec) are called {\it hyperparameters}. In addition to providing a
method for model selection, \rev{by integrating over parameters to
  calculate the evidence rather than using a ``best'' point estimate
  of the parameters, ME avoids the ill-posedness problem associated
  with ML.}

Although ME provides an approach to model selection, calculation of
the evidence does not, on its own, provide an estimate for \thv. The
VBEM approach to estimating evidence does, however,
provide a mechanism to estimate \thv. \rev{VBEM can be thought of as
  an extension of EM for ME. Both the VBEM algorithm and considerations for
  choosing priors are discussed in Methods.}

%% file: fret_background.tex
Before building a GM describing smFRET data, it is helpful to review
briefly the experimental method. The experimental technique is based
on the spectroscopic phenomenon that, if the emission spectrum of a
polar chromophore (donor) overlaps with the absorption spectrum of
another polar chromophore (acceptor), electromagnetic excitation of
the donor can induce a transfer of energy to the acceptor via a
non-radiative, dipole-dipole coupling process termed florescence
resonance energy transfer (FRET) \cite{ForsterADP48}. The transfer
efficiency between donor and acceptor scales with the distance between
molecules (r) as $1/r^6$, with FRET efficiencies most sensitive to r
in the range of $1-10$nm. Because of this extraordinary sensitivity to
distance, FRET efficiency can serve as a molecular ruler, allowing an
experimentalist to measure the separation between donor and acceptor
by stimulating the donor with light and measuring emission intensities
of both the donor ($I_D$) and acceptor ($I_A$)
\cite{HauglandPNAS67}. Usually a summary statistic called the ``FRET
ratio'' is used to report on
molecular distance rather than the ``raw'', 2-channel $\{I_A$,$I_D\}$
data, although inference of the raw 2-channel data is possible as well
\cite{WigginsBJ09}. The FRET ratio is given by
\begin{equation}
\FRET\ = \frac{I_A}{I_D + I_A}.
\label{eq:fret_ratio}
\end{equation}

When the donor and acceptor are attached to an individual protein,
nucleic acid, or other molecular complex, the FRET signal can be used
to report on the dynamics of the molecule to which the donor and
acceptor are attached (see \fig{fig:FRET}). When the experiment is
crafted to monitor individual molecules rather than ensembles of
molecules, the process is termed single molecule FRET (smFRET). For
many biological studies, such as the identification and
characterization of the structural dynamics of a biomolecule, smFRET
must be used rather than FRET; the majority of molecular
dynamics cannot be observed from ensemble averages. Often the molecule
of interest adopts a series of locally stable conformations during a
smFRET time series. From these data, the experimentalist would like to
learn (1) the number of locally stable conformations in the data (i.e.
states) and (2) the transition rates between states. Although it is
theoretically possible use the FRET signal to quantify the distance
between parts of a molecule during a time series, there are usually
too many variables affecting FRET efficiency for this to be practical
\cite{EatonPNAS05}. Consequently, smFRET is usually used to extract
quantitative information about kinetics (i.e. rate constants) but only
qualitative information about distances.

The photophysics of FRET have been studied for over half a century, but
the first smFRET experiments were only carried out about fifteen years
ago \cite{WeissPNAS96}. The field has been growing exponentially
since, and hundreds of smFRET papers are published annually 
\cite{HaARB08}. Diverse topics such as protein folding
\cite{WeissPNAS00}, RNA structural dynamics \cite{ChuSCI02}, and
DNA-protein interactions \cite{HaNAT09} have been investigated via
smFRET. The size and complexity of smFRET experiments
has grown substantially since the original smFRET publication.
A modern smFRET experiment can generate thousands of time series to be
analyzed \cite{GonzalezPNAS09}.

%% file: modeling_fret.tex
A model of the smFRET time series for a molecule transitioning between
a series of locally stable conformations should capture several
important aspects of the process \cite{TalagaJPC03}. The observable
smFRET signal is a function of the hidden conformation of the
molecule. The noise of each smFRET state can be assumed to be
Gaussian, and the hidden conformations are assumed to
be discrete and finite in number. The probability of transitioning to
a new molecular conformation should be a function of the current
conformation of the molecule (\eg, the DNA in \fig{fig:FRET} is more
likely to be zipped at time $t+1$ if it is zipped at time $t$). The
CCD cameras commonly used in smFRET experiments naturally bin the
data temporally, so it is convenient to work with a model where time
is discrete. The GM expressing these features is called a hidden
Markov model (HMM) and is shown in \fig{fig:hmm_gm_all}A. From the graph,
it can be seen that the probability of the observed and latent
variables factorizes as
\begin{equation}
p(\D,\Z|\thv,\K) =
p(\z_1|\thv,\K)\left[\prod_{t=2}^Tp(\z_t|\z_{t-1},\thv,\K) \right]
\prod_{t=1}^Tp(\d_t|\z_t,\thv,\K).
\label{eq:HMM}
\end{equation}

Here, \thv\ must include parameters for the probability that the time
series begins in each state ($p(\z_1=k) \equiv \pi_k$); parameters for
transition probabilities between states ($p(\z_{t+1}=k|\z_t=j) \equiv
\ajk$); and parameters for the noise of the emissions of each state
($p(\d_t|z_t=k) = \mathcal{N}(\d_t|\mu_k,\lambda_k)$, where $\mu_k$
and $\lambda_k$ are the mean and precision of the Gaussian). It is
necessary to model $p(\z_1)$ separately from all other transition
probabilities since it is the only hidden state probability which does
not depend on $z_{t-1}$. The \ajk\ are commonly represented as a
matrix, \A, called a transition matrix. The probability the time
series begins in the \kth\ state and transition probabilities between
states are drawn from multinomial distributions defined by \pivec\ and
the rows of \A, respectively.  The GM for this HMM is shown in
\fig{fig:hmm_gm_all}B. From the GM it can be seen that
\begin{eqnarray}
  \jointlike \prior &=&
  p(\z_1|\pivec \kopt)\left[\prod_{t=2}^Tp(\z_t|\z_{t-1}, A \kopt) \right]
  \prod_{t=1}^Tp(\d_t|\z_t,\vec{\mu},\vec{\lambda} \kopt) \times \nonumber \\
  & & p(\pivec|u_{\pi} \kopt)p(A|\vec{u}_A \kopt)
  p(\vec{\mu}|\vec{u}_m,\vec{u}_{\beta},\vec{\lambda} \kopt)
  p(\vec{\lambda}|\vec{u}_a,\vec{u}_b \kopt). 
\label{eq:hmm_bayes}
\end{eqnarray}

\rev{For a time series of length T where each latent variable can take
  on K states, a brute summation over all possible states requires
  $O(\K^T)$ calculations.  By exploiting efficiencies in the GM and
  using the sum-product algorithm, this summation can be performed
  using $O(\K^2T)$ calculations (which can be seen by noting that the
  latent state probabilities in \eq{eq:hmm_bayes} factorize into
  $p(\z_t|\z_{t-1},A,\K)$, where each of the $T$ latent states has
  $\K^2$ possible combinations of states).} The sum-product algorithm
applied to the HMM is called the forward-backward algorithm or the
Baum-Welch algorithm \cite{rabiner89}, and the most probable
trajectory is called the Viterbi path \cite{viterbi}.

\rev{There are several assumptions of this model which should be
  considered. First, although it is common to assume the noise of
  smFRET states is Gaussian, the assumption does not have a
  theoretical justification (and since FRET intensities can only be on
  the interval $(0,1)$, and the Gaussian distribution has suport
  $(-\infty,\infty)$, the data cannot be truly Gaussian). Despite this
  caveat, several groups have successfully modeled smFRET the data as
  having Gaussian states \cite{SachsPRSL97, HaBJ06, WigginsBJ09}. We
  note that other distributions have been considered as well
  \cite{HAJPCB10}.

    Second, the HMM assumes that the molecule instantly switches between
    hidden states. If the time it takes the molecule to transition between
    conformations is on the same (or similar) order of magnitude as
    the time it spends within a conformation, the HMM is not an
    appropriate model for the process and a different GM will
    be needed. For many molecular processes, such as protein domain
    rearrangements, the molecule transitions between conformations
    orders of magnitude faster than it remains in a conformation and
    the HMM can model the process well \cite{Creighton92}.

    Third, the HMM is ``memoryless'' in the sense that, given its
    current state, the transition probabilities are independent of the
    past. It is still possible to model a molecule which sometimes
    transitions between states quickly and sometimes transitions
    between states slowly (if, for example, binding of another small
    molecule to the molecule being studied changes its transition
    rates \cite{GonzalezPNAS09}). This situation can be modeled using
    two latent states for each smFRET state. The two latent states
    will have the same emissions model parameters, but different transition
    rates.}

%% file: validation.tex
A software package, vbFRET, implementing the VBEM algorithm for this
HMM was written and described in \cite{WigginsBJ09}, along with an
assessment of the algorithm's performance on real and synthetic data.
An illustration of the method is shown here, demonstrating three of
its most important abilities: the ability to perform model selection;
the ability to learn posterior parameter distributions; and the
ability to idealize a time series. These abilities are demonstrated on
three synthetic $\K=3$ state time series, shown in
\fig{fig:illustration}C. The traces all have $\mu = \{0.25, 0.5,
0.75\}$ and identical hidden state trajectories. The noise of each
hidden state is $\sigma = 0.015$ for trace 1 (unrealistically
noiseless), $\sigma = 0.09$ for trace 2 (a level of noise commonly
encountered in experiments), and $\sigma = 0.15$ for trace 3
(unrealistically noisy).

{\bf Model selection:} For each trace, $\lb$, the lower bound of the
log(evidence), was calculated for $1 \geq \K \geq 7$. The results are shown in
\fig{fig:illustration}A, with the largest value of $\lb$ for each
trace shown in red. For traces 1 and 2, $\lb$ peaks for $\Ks =3$,
correctly inferring the complexity of the model. For trace 3, the
noise of the system is too large, given the length of the trace, to
infer three clearly resolved states. For this trace $\lb$ peaks at
$\Ks=2$. This result illustrates an important consideration of
evidence based model selection: states which are distinct in a
generative model (or an experiment generating data) may not give rise
to statistically significant states in the data generated. For
example, two states which have identical means, variances, and
transition rates would be statistically indistinguishable from a
single state with those parameters. When states are resolvable,
however, ME-based model selection is generally effective, as
demonstrated in traces 1 and 2.

{\bf Posterior distributions:} The ability to learn a complete
posterior distribution for $\thv$ provides more information than
simply learning an estimate for $\thv$, and is a feature unique to
Bayesian statistics. The maximum of the distribution, denoted \thvmap,
can be used as an estimate of \thv\ (e.g., if idealized trajectories
are needed). The subscript here differentiates it from the estimate in
the absence of the prior, \thvml. The curvature of the distribution
describes the certainty of the \thvmap\ estimate.  As a demonstration,
the posterior for the mean of the lowest smFRET state of each trace is
shown in \fig{fig:illustration}B. The X and Y axes are the same in all
three plots, so the distributions can be compared. As expected, the
lower the noise in the trace, the narrower the posterior distribution
and the higher the confidence of the estimate for $\mu$. The estimate
of $\mu$ for trace 3 is larger than in the other traces because
$\Ks=2$; some the middle smFRET state data are misclassified as
belonging to the low smFRET state.

{\bf Idealized trajectories:} Idealized smFRET trajectories can be a
useful visual aid to report on inference. They are also a
necessity for some forms of post-processing commonly used at present,
such as dwell-time analysis \cite{GonzalezPNAS09}. Idealized
trajectories can be generated from the posterior learned from VBEM by
using \thvmap\ to calculate the most probable hidden state trajectory
(the Viterbi path) \cite{viterbi}. The idealized trajectories for each
trace are shown in \fig{fig:illustration}C.  For traces 1 and 2, where
$\Ks$ is correctly identified, the idealized trajectory captures the
true hidden state trajectory perfectly.  Because of the model
selection and well-posedness of ME/VBEM the idealized trajectories
learned with this method can be substantially more accurate than those
learned by ML for some data sets \cite{WigginsBJ09}.

%% file: conclusion.tex
This manuscript demonstrates how graphical modeling, in conjunction
with a detailed description of a biophysical process, can be used to
model biophysical time series data effectively. The GM designed here
is able to model smFRET data and learn both the number of states in
the data and the posterior parameter values for those states. The
ME/VBEM methodology used here offers several advantages over the more
commonly used ML/EM inference approach, including
intrinsic model selection and a well-posed optimization. All modeling
assumptions are readily apparent from the GM. 
The GM framework with inference using ME/VBEM is highly flexible
modeling approach which we anticipate will be applicable to a wide
array of problems in biophysics.

%% file: vbem.tex
Unfortunately, calculation of \eq{eq:me} requires a sum over all K
settings for each of T extensive variables \Z\ (where T is the length
of the time series). Such a calculation is numerically intractable,
even for reasonably small systems (\eg, K=2, T=100) so an
approximation to the evidence must be used. Several approximation
methods exist, such as Monte Carlo techniques, for numerically
approximating such sums \cite{neal1993piu}. The method we will
consider here is VBEM.

One motivation for the VBEM algorithm is the following simple
algebraic identity \cite{Bishop06}. Since Bayesian analysis treats
latent variables (\Z) and unknown parameters (\thv) the same way this
section will lump them both into \X\ for notational simplicity. Let
$q(\X)$ be any probability distribution \rev{over \X}. Then,
\begin{eqnarray}
\log \ev & = & \int q(\X) \log\left(\ev\right) d\X\ \label{eq:lb1} \\
& = & \int q(\X) \log\left(\frac{p(\D,\X|\rev{\uvec},\K)}{p(\X|\D,\rev{\uvec},\K)}
 \right) d\X\  \label{eq:lb2} \\
 & = & \int q(\X) \log\left(\frac{p(\D,\X|\rev{\uvec},\K)q(\X)}{p(\Z|\D,\rev{\uvec},\K)q(\X)}
 \right) d\X\  \label{eq:lb3}\\
 & = & \int q(\X) \log\left(\frac{p(\D,\X|\rev{\uvec},\K)}{q(\X)}
 \right) d\X\ \nonumber \\
& &  -\int q(\X) \log\left(\frac{p(\X|\D,\rev{\uvec},\K)}{q(\X)}
 \right) d\X\ \label{eq:lb4}\\
&=& \lb + \Dkl{\qzt}{\jointpost} \label{eq:lb5}.
\end{eqnarray}
Summations over the discrete components of \X\ should be included in
these equations, but are omitted for notational simplicity.  The
equality in \eq{eq:lb1} results from the requirement that $q(\X)$ be a
normalized probability; \eq{eq:lb2} rewrites \ev\ in terms of a
conditional probability; and \eq{eq:lb5} reinserts $\{\Z,\thv\}$ for
$\X$ and renames the two terms in \eq{eq:lb4} as $\lb$, the lower
bound of the log(evidence), and the Kullback–Leibler divergence,
respectively.

Using Jensen's inequality, it can be shown that 
\begin{equation}
  \Dkl{q}{p} \geq 0,
\label{eq:dkl_ineq}
  \end{equation}
  with equality when $q=p$. Consequently,
\begin{equation}
\log\left(\ev\right) \geq \lb,
\end{equation}
\ie, $\exp(\lb)$ is a lower bound on the model's
evidence. \eq{eq:dkl_ineq} implies that \lb\ is maximized when $\qzt$
is equal to \jointpost. As a corollary, from this it follows that
$q(\theta)$ approximates $\post$, the {\it posterior} distribution of
the parameters.  Therefore, the optimization simultaneously performs
model selection (by finding a \K\ which maximizes \ev) and inference
(by approximating \jointpost).

The approach suggested by Eqs.~$\ref{eq:lb1}$--$\ref{eq:dkl_ineq}$ is to
replace an intractable calculation with a tractable bound
optimization. If \like\ is in the exponential family and a conjugate
prior is used, then the only assumption about $\qzt$ needed is that
$q(\Z,\thv) = q(\Z)q(\thv)$ (\ie, it factorizes into a function of \Z\
and a function of \thv) for the inference problem to be tractable
using VBEM \cite{GhahramaniBA06}. In addition, under these conditions
\post\ will have the same functional form as \prior. The VBEM
algorithm is similar to EM, but rather than iteratively using guesses
for \thvml\ to set \Z\ and guesses for \Z\ to set \thvml\, the update
equations iterate between \cite{BealPHD03, Bishop06}:
\begin{equation}
\mathrm{VBE:\ } q(\Z) = \frac{1}{\mathcal{Z_\Z}} \exp \left(
\mathbb{E}_{q(\thv)}\left[\log\left(p(\D,\Z|\thv,\K)\prior\right)\right] \right)
\label{eq:update_z}
\end{equation}
\begin{equation}
\mathrm{VBM:\ } q(\thv) = \frac{1}{\mathcal{Z_{\thv}}} \exp \left(
\mathbb{E}_{q(\Z)}\left[\log\left(p(\D,\Z|\thv,\K)\prior\right)\right]
\right).
\label{eq:update_thv}
\end{equation}
Here $\mathbb{E}$ denotes the expected value with respect to the
subscripted distribution and $\mathcal{Z}$ is a normalization
constant. Whereas the log(\ev) is a log of a sum/integral, the right
hand sides of Eqs.~\ref{eq:update_z} \& \ref{eq:update_thv} both involve
the sum/integral of a log. This difference renders $\log(\ev)$
intractable, yet Eqs.~\ref{eq:update_z} \& \ref{eq:update_thv}
tractable.

An interesting and potentially useful feature of the q(\thv) learned
from VBEM is that when \K\ is chosen to be larger than the number of
states supported by the data, the optimization leaves the extra states
unpopulated. This propensity to leave unnecessary states unpopulated
in the posterior, sometimes called ``extinguishing'', is a second form
of model selection intrinsic to VBEM, which is in addition to the
model selection described by \eq{eq:me}.  An explanation for this
behavior can be found in Chapter 3 of \cite{Bishop06}.

%% file: priors.tex
Several considerations should go into choosing a prior. Choosing
distributions which are conjugate to the parameters of the likelihood
can greatly simplify inference \cite{GhahramaniBA06}. Priors can be
chosen to minimize their influence on the inference. Such priors are
called ``weak'' or uninformative. Alternatively, priors can also be
chosen to respect previously obtained experimental observations
\cite{BealPHD03}. It is important to check that inference results do
not heavily depend on the prior (\eg\ doubling or halving
hyperparameter values should not affect inference results).

The conjugate prior of a multinomial distribution is a Dirichlet
distribution: $p(\pi_{1},\ldots,\pi_{K})$ = Dir($\pivec|u_{\pi}$);
$p(a_{k1},\ldots,a_{kK})$ = Dir($a_{k,1-K}|u_A^k$). Expressed in terms
of precision $\lambda$, rather than variance $\sigma^2$ (where
$\lambda$ = $1/\sigma^2$), the conjugate prior for the
mean and precision of a Gaussian is a Gaussian-Gamma distribution:
$p(\mu_k,\lambda_k)$ =
$\mathcal{N}(\mu_k|u_m^k,(u_\beta^k\lambda_k)^{-1})$Gam($\lambda_k|u_a^k,u_b^k$).

Here, hyperparameters were set so as to give distributions consistent
with experimental data and to influence the inference as weakly as
possible: $u_{\pi}^k = 1$, $u_a^{jk} = 1$, $u_{\beta}^k = 0.25$,
$u_m^k = 0.5$, $u_a^k = 2.5$ and $u_b^k = 0.01$, for all values of
$k$.  Qualitatively, these hyperparameters specify probability
distributions over the hidden states such that it is most probable
that the hidden states are equally likely to be occupied and equally
likely to be transitioned to.  Quantitatively, they yield $\langle
\mu_k \rangle= 0.5$ and mode$[\sigma] \approx 0.08$, consistent with
experimental observation:
\begin{equation}  \frac{1}{\sqrt{{\rm
    mode}[\lambda_k]}} = \sqrt{\frac{u_b^k}{(u_a^k-1)}} \approx 0.08 ~\forall k.
\end{equation}

%% file: supporting.tex
Synthetic traces were generated in \matlab\ using 1-D Gaussian noise
for each hidden state and a manually determined hidden state
trajectory. All traces were analyzed by vbFRET
\cite{WigginsBJ09}, using its default parameter settings, for $1 \geq
\K  \geq 7$, with 25 random restarts for each value of \K. The restart with
the highest evidence was used to generate the data in
\fig{fig:illustration}. The posterior probability of $\mu_k$ is given
by $\mathcal{N}(\mu_k|v_m^k,(v_\beta^k\lambda_k)^{-1})$, where
$\vec{v}$ are the hyperparameters of the posterior. The data in
\fig{fig:illustration}B were generated using this equation with
$\lambda_k$ fixed at its most probable posterior value.

%% file: captions_for_figures.tex
\clearpage
\section*{Figures}
   
\begin{figure}[h!]
\begin{flushleft}
    \includegraphics{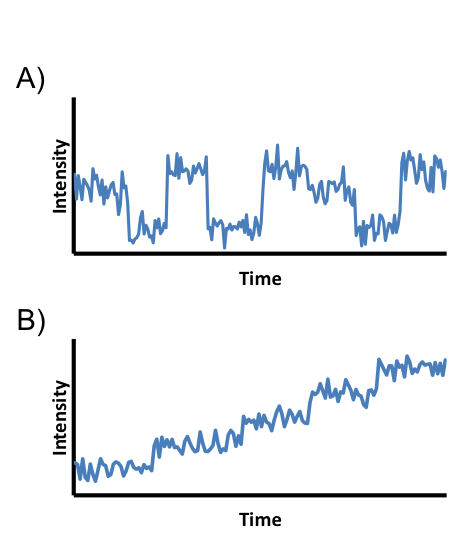}
    \caption{Examples of types of commonly encountered biophysical
      time series data. (A) A time series for a molecule
      transitioning between a series of locally stable
      conformations. Such data often arise, for example, when studying
      protein domain movements or the dynamics of polymers tethered to
      a surface. (B) A time series for a molecule undergoing a
      stepping process. Such data often arise, for example, when
      studying proteins with unidirectional movements, e.g., helicases
      and motor proteins.}
   \label{fig:time_series}
\end{flushleft}
\end{figure}

\clearpage

\begin{figure}[h!]
\begin{flushleft}
    \includegraphics{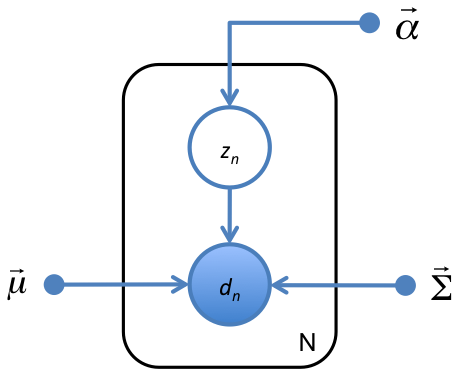}
    \caption{A GM for the problem of learning genders of boys and
      girls from a table of their heights and weights. The gender of
      the \nth\ child is denoted $\z_n$. The 2-dimensional vector of
      the child's height and weight is denoted $\d_n$ The mean hight
      and weight for each gender, variances of height and weight for
      each gender, and probability of belonging to each gender are
      denoted by $\muvec$, $\vec{\Sigma}$, and $\vec{\alpha}$,
      respectively.  Observed variables are represented by open
      circles, hidden variables are represented by filled circles, and
      fixed parameters are represented by dots. To avoid drawing
      nodes for all $N$ hidden and observed variables, the variables
      are shown once and placed inside a plate which denotes the
      number or repetitions in the lower right corner. This GM
      specifies the conditional factorization of $p(\D,\Z,\thv)$ shown
      in \eq{eq:boy_girl}.}
   \label{fig:gm_example}
\end{flushleft}
\end{figure}

\begin{figure}[h!]
\begin{flushleft}
    \includegraphics{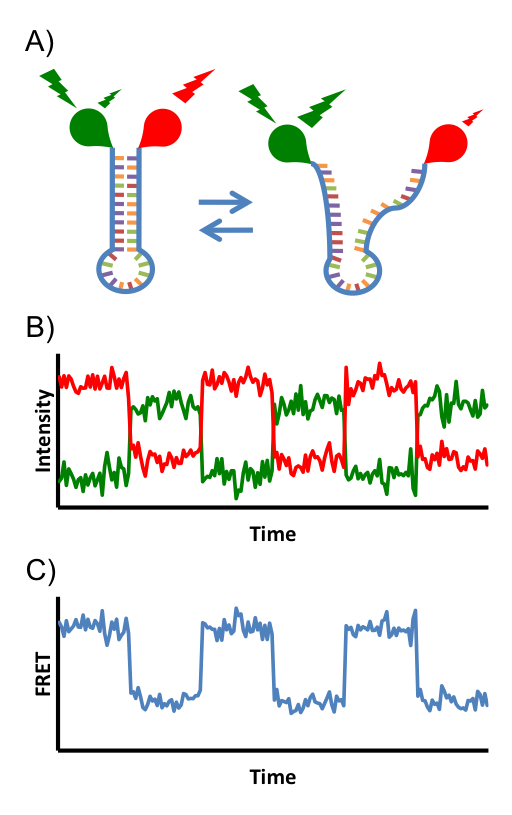}
    \caption{ {(A)} Cartoon of a smFRET experiment studying the
      zipping/unzipping of a DNA hairpin. A FRET donor chromophore
      (green balloon) and acceptor chromophore (red balloon) are
      attached to the DNA. When the DNA is zipped (left), exciting the
      donor with green light causes the majority of energy to be
      transferred to the acceptor. The donor will fluoresce dimly and
      the acceptor will fluoresce brightly. When the DNA is unzipped,
      the probes are too far apart for efficient FRET. Exciting the
      donor in this conformation causes it to fluoresce brightly and
      the acceptor to fluoresce dimly. {(B)} The two channel
      (donor, acceptor) time series generated by the DNA as it
      transitions between zipped (bright red, dim green) and unzipped
      (dim red, bright green). {(C)} The 1D FRET transformation of the
      time series from B, calculated with \eq{eq:fret_ratio}. The
      closer the probes, the more intense the signal. Time series of
      this summary statistic are commonly used for analysis.}
   \label{fig:FRET}
\end{flushleft}
\end{figure}

\clearpage
\begin{figure}[h]
\begin{flushleft}
    \includegraphics{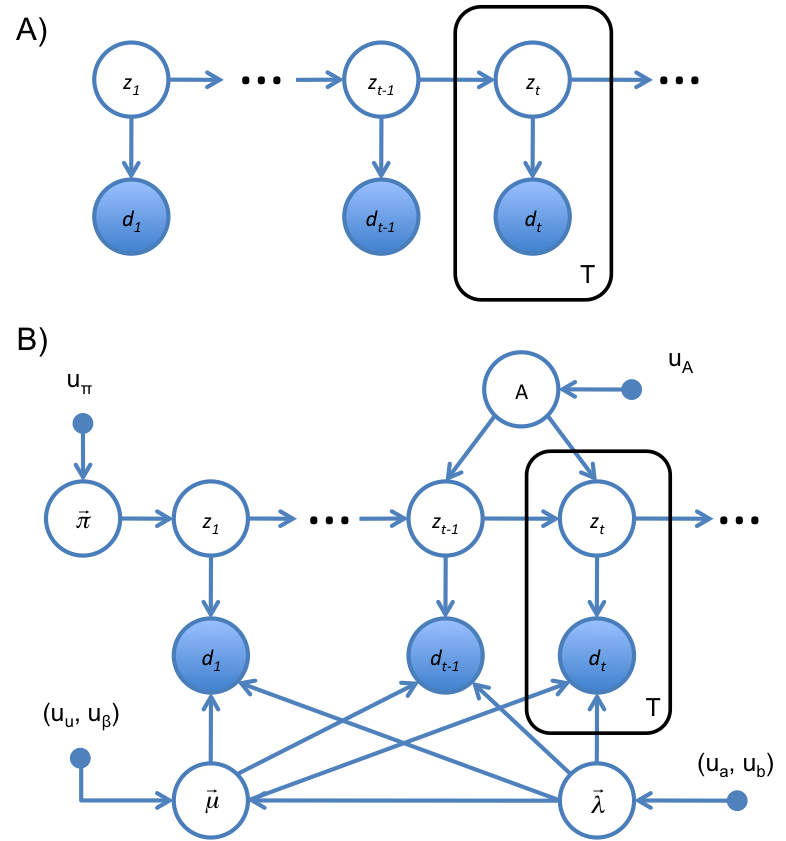}
    \caption{(A) The HMM as a GM. At each time step, $t$, the system
      occupies a hidden state, $z_t$ and produces an observable
      emission, $d_t$, drawn from $p(\d_t|z_t)$. In turn, $\z_t$ is
      drawn from $p(\z_t|\z_{t-1})$. (B) Complete GM for the HMM used to
      describe smFRET data in this work. Following the Bayesian
      treatment of probability, all unknown parameters are treated as
      hidden variables, and represented as open
      circles. 
      Emissions are assumed to be Gaussian, with mean $\muvec$ and
      precision $\lambdavec$. Transition rates are multinomial, with
      probabilities given by \A. The probability of initially
      occupying each hidden state is multinomial as well, with
      probabilities given by $\pivec$. Equations for these
      distributions are described in the text below \eq{eq:HMM}. This
      GM specifies the conditional factorization of $p(\D,\Z,\thv)$
      shown in \eq{eq:hmm_bayes}.}
   \label{fig:hmm_gm_all}
\end{flushleft}
\end{figure}

\clearpage
\begin{figure}[h]
\begin{flushleft}
    \includegraphics{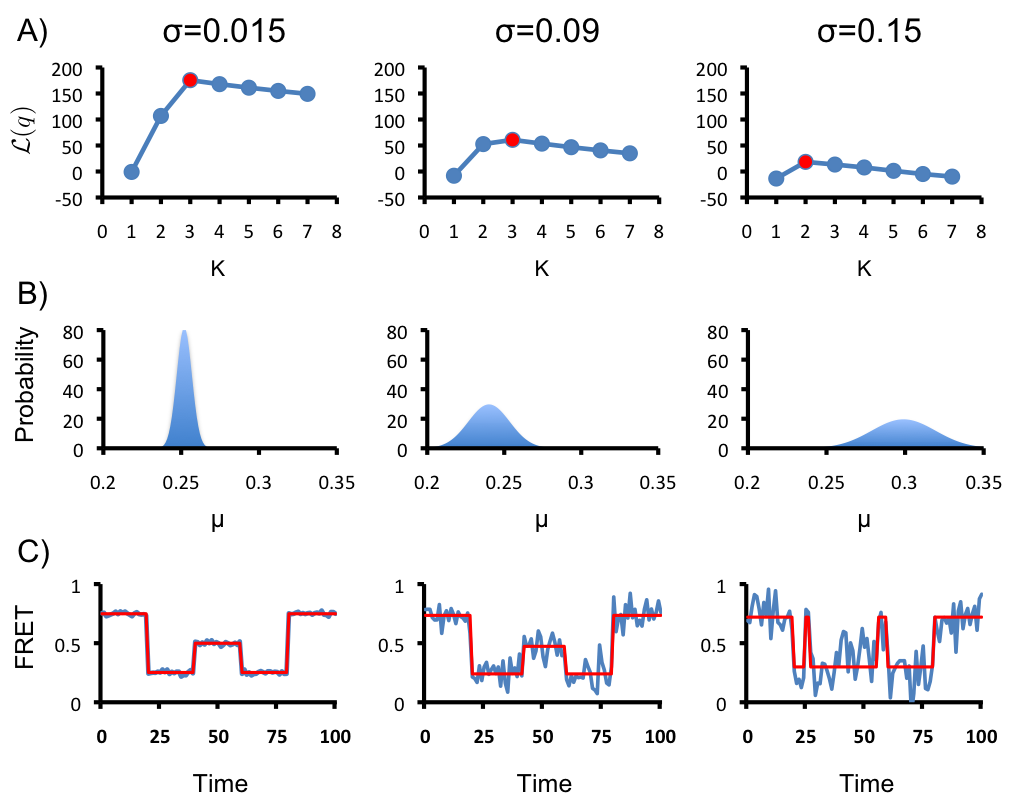}
    \caption{(A) Model selection using ME. Inference using $1 \leq \K
      \leq 7$ hidden states was performed for each trace. The results
      with the highest $\lb$ are shown in red. (B) The posterior
      parameter distribution for the lowest-valued smFRET state
      inferred in each time series. The width of the posterior
      increases with the noise of the smFRET states, indicating lower
      confidence in the parameters learned from inference on noisier
      time series. (C) The idealized trajectories (red) inferred for
      each time series (blue) using the most probable parameters of the
      inference with the highest $\lb$.}
   \label{fig:illustration}
\end{flushleft}
\end{figure}